\documentclass[useAMS,usenatbib]{mn2e}
\input psfig.tex

\title[LFs with rigour]{Rigorous luminosity function determination in presence 
of a background: theory and application to 
two intermediate redshift clusters\thanks{Based on observations
collected at the European Southern Observatory, Chile,
and, in part, on observations with the NASA/ESA Hubble Space Telescope}
}
\author[Andreon, et al.]{S. Andreon,$^1$\thanks{andreon@brera.mi.astro.it}  
G. Punzi,$^2$ and A. Grado,$^3$ \\
$^1$INAF--Osservatorio Astronomico di Brera, Milano, Italy\\
$^2$INFN \& Scuola Normale Superiore, Pisa, Italy\\
$^3$INAF--Osservatorio Astronomico di Capodimonte, Napoli, Italy\\
}
\date{Accepted ... Received ... MF50rv}
\pagerange{\pageref{firstpage}--\pageref{lastpage}}
\pubyear{2004}
\begin{document}

\maketitle

\label{firstpage}

\begin{abstract}  In this paper we present a rigorous derivation of the
luminosity function (LF) in presence of a background. Our approach  is
free from the logical contradictions of assigning negative values to
positively defined quantities and avoid the use of incorrect estimates for
the 68 \% confidence interval (error bar). It accounts for Poisson
fluctuations ignored in previous approaches and does not requires binning
of the data. The method is extensible to more complex situations, does not
require the existence of an environment--independent LF, and clarifies
issues common to {\it field} LF derivations.  We apply the method to two
clusters of galaxies at intermediate redshift ($z\sim0.3$) with among the
deepest and widest $K_s$ observations ever taken. Finally, we  point out
short-comings of flip--flopping magnitudes.
\end{abstract}

\begin{keywords}  
Galaxies:
evolution --- galaxies: clusters: general --- galaxies: clusters: --- method:
statistic
\end{keywords}

\section{Introduction}

The luminosity function (LF), i.e. the number of galaxies per unit luminosity
and volume is one of the fundamental quantities of observational cosmology:
it is interesting in its own, and it is a necessary ingredient (weight) in
most cosmological measures dealing with galaxies. The history of the LF
determination dates back to Zwicky (1957) at least. This debate with
Hubble (Zwicky 1951, Hubble 1936) around the shape of the LF is  
one of the pillar of the history of the LF determination.

With the advent of large surveys, such as 2dF (Folkes et al. 1999), 
SDSS (York et al. 2000) and of the Virtual Observatory, samples grow by
orders of magnitude, and it is nowadays common to deal with more than
one thousand galaxies when computing the LF. However, at the extremes
of absolute magnitude ranges or in special environments or for
certain galaxy types, the number of
galaxies is often low. Methods used for the LF computation also improved
along the years (see citations in sec 3).

In Andreon (2004) we showed how much the neglected observer prior 
influences the
found result (error and confidence interval, an example along the same line is
presented in Blanton et al. 2003). This paper has a twofold
aim: improve the method in the LF determination and apply it 
to the best data (useful for the LF determination) ever taken in
$K$-band.

\medskip
The paper is organized as follows: in section 2 the data and the data 
reduction are presented. In Section~3, new statistical method is
presented. In Section~4 we derive the LF.
The discussion and a summary are presented in Section~5.

We assume $H_0=70$ km s$^{-1}$ Mpc$^{-1}$ 
$\Omega_\Lambda=0.7$, $\Omega_{M}=0.3$.

\begin{figure*}
\caption[h]{The $K_s$ band image of AC\,114. The field of view is
$\sim5\times5$ arcmin. North is up and East is to the left.}
\end{figure*}

\section{Data and data reduction}

\subsection{AC\,114 \& AC\,118}

AC\,114 and AC\,118 are among the most 
observed clusters at intermediate redshift. Discovered by 
Couch \& Newell (1984) and later by Abell, Corwin \& Olowin (1989),
they have been the focus of extensive studies:
spectroscopic observations (e.g. Couch \& Sharples 1987), near
infrared imaging (e.g. Barger et al. 1996), {\it Hubble Space Telescope}
observations (e.g. Couch et al. 1998), mass determination 
through gravitational lensing experiments  (Smail et al. 1997), 
galaxy evolution studies (Barger et al. 1996; Couch et al. 1998; Jones, Smail
\& Couch 2000; Couch et al. 2001), etc. 
AC\,114 is a regular massive cluster,
whereas AC\,118 is a massive merging system. A detailed description of
these two clusters may be found in the mentioned papers.

AC\,114 observations were carried out at the 3.5 m NTT with SOFI (Moorwood, Cuby 
and Lidman, 1998) for four nights during fall 1998. SOFI is equipped with a 
$1024\times1024$ pixel Rockwell ``Hawaii" array. In its large field mode 
the pixel size is 0.292 arcsec and the field of view $5\times5$ arcmin. 
The field was observed in the near--infrared $K_s$ passband ($\lambda_c=2.2 
\mu$; $\Delta\lambda \sim 0.3 \mu$) during four photometric nights with
good seeing ($FWHM<0.8$ arcsec). The total useful exposure time is 18840 s, 
resulting from the coaddition of many short jittered exposures.  
Photometric calibration has been obtained by observing a few standard 
stars, interspersed with AC\,114 observations, taken from the 
list of Infrared NICMOS Standard Stars published in Persson et al. (1998). 
Fig. 1 shows the final $K_s$ image of AC\,114. This image has a seeing of
0.8 arcsec.

AC\,118 observations were carried out with the same instrument, the night
after AC\,114 observations, and are fully described in Andreon (2001, Paper I).

All images have been reduced as in paper I. Shortly, they are
flat--fielded by flaton--flatoff. In order to test the accuracy of the 
flat--fielding, a standard 
star has been observed in 8 chip locations. The root mean square
variation of his magnitude is 0.008 mag.
Since the RMS deviation is small, our images do not require a 
supplementary illumination correction.  The background has been
removed by using Eclipse (Devillard, 1997), taking advantage of
the telescope nodding during the observations. Images have been 
combined using the task {\tt imcombine} under IRAF using integer
pixel shifting.

\begin{table*}
\caption{The data}
\begin{tabular}{lllll}
\hline
& AC\,114 & AC\,118 & CDF-S & HDF-S \\
\hline
Exposure time (min)	      & 314  & 265  & 80--180   & 180--300    \\
Seeing	(FWHM, arcsec)	      & 0.73 & 0.75 & $\sim$0.8 & 0.9-1.0     \\
Complet. mag ($\phi=4.^"4$)   & 20.3 & 20.5 & 19.5-20.0 & 20.25	     \\
Area (arcmin$^2$)	      & 23.7 & 23.7 & 242.0-45.0 & 47.0	      \\
\hline
\end{tabular}
\end{table*}

\begin{figure*}
\psfig{figure=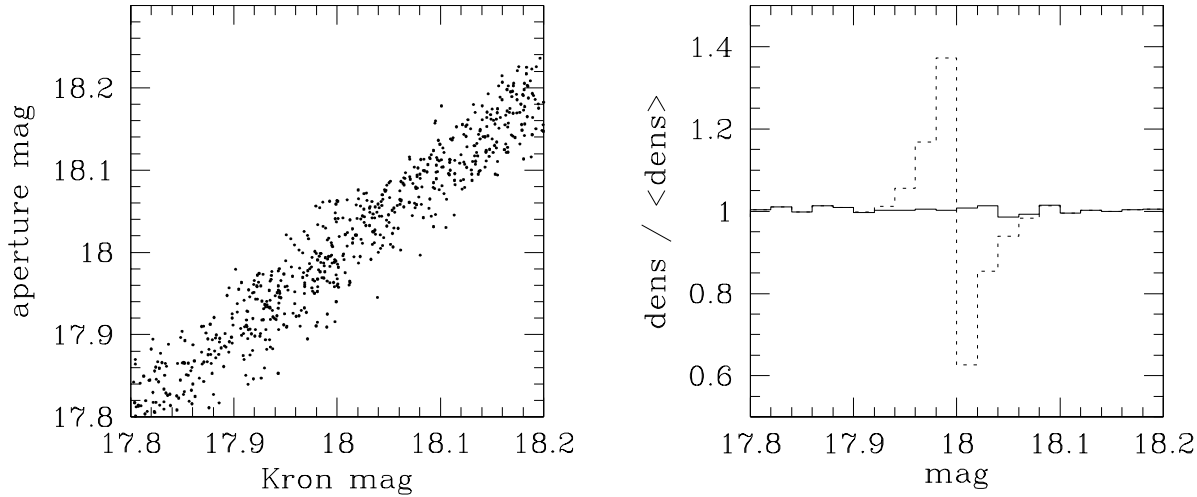,width=16truecm}
\caption[h]{The magnitude bias in a Monte Carlo simulation with Gaussian noise
($\sigma=0.03$ mag) (shown in the left panel), adopting an orthogonal (solid
histogram)  or vertical cut (dotted histogram).} \end{figure*}

\subsection{Control field: CDF-S \& HDF-S}

As control field we use the  Hubble Deep Field South 1 and 2 (HDF-S) images
(Da Costa et al. 1998), already used for AC\,118, and therein described,
supplemented by Chandra Deep Field South (CDF--S, hereafter) images (Vandame
et al., 2001; 2004 and Rengelink et al. 1998). We only remind that all
these images have been taken with the same instrument, filter and telescope as
the AC\,114 \& AC\,118 images, that cluster images are interspersed to
control field images, hence ensuring an almost perfect homogeneity 
of the data. The basic data reduction of control and science fields is based
on the same software (Eclipse). Two major differences occurs: science
data have not been resampled, in order not to correlate the noise of
adjacent pixels, and science data are combined with more attention to
flux (see paper I for details), allowing us to claim a better photometric
calibration precision for cluster images (better than 0.01 mag) than other
authors claim for the control field (around 0.05 mag).

The 16 SOFI pointings of the CDF-S guarantee a large area coverage (242
armin$^2$) down to $K_s=19.5$ and hence a good determination of the galaxy
counts in the control field. Three of them, covering 45 arcmin$^2$, are
exposed longer and reach $K_s=20$, hence supplementing the 47 arcmin$^2$ of
the HDF-S down to $K_s=20.25$ used in Paper I. At $K_s<18$ HDF-S shows a
marginally high overdensity with respect to CDF-S. Therefore, we arbitrary
remove the bright part of the HDF--S galaxy counts (that in any case carry
a negligible weight, given the small observed area of the HDF--S).

Table 1 shows a summary of all observations. The area coverage of
the CDF--S alone is larger than the latest published galaxy
counts (Crist{\' o}bal-Hornillos et al. 2003), down
to their completeness limit ($K\sim19.5$). 

\subsection{Photometry and flip--flopping magnitudes}

Objects has been detected by using SExtractor version 2.2 (Bertin \& Arnouts 1996).
For AC\,114 we made use of the RMS map for a clean detection, as we did for
AC\,118. Due to the varying exposure time across the field of each image,
due to the dithering, we consider here only the central square areas listed
in Table 1. 

Galaxy are extended objects, hence their luminosity depends on way their
border are defined. We improve our magnitude definition with respect to
paper I: here we adopt Kron magnitudes (see Kron 1980 for the exact
definition, and Bertin \& Arnouts 1996 for a software implementation) for
bright ($(K_{Kron}+K_{aper})/2<18$ mag) objects and aperture (in a 4.4
arcsec aperture) otherwise. In paper I the cut were performed along one of
the axis ($K_{Kron}$), and not orthogonally to the $K_{Kron}$ {\it vs}
$K_{aper}$ relationship, spuriously producing a density variation 
in un--binned distribution of galaxy counts (not actually
used in that paper, but used here) due to the spread around the $K_{Kron}$
{\it vs} $K_{aper}$  relationship. A Monte Carlo simulation of what occurs is shown
in Fig 2. On the left we show a linear relationship between the Kron and
aperture magnitude, with a Gaussian small scatter ($\sigma=0.03$ mag) and no
bias. On the right, we histogram
the galaxy counts with a cut orthogonal to the Kron vs aperture mag
relationship (solid histogram), and at a fixed Kron magnitude (dotted
histogram). The latter histogram presents a huge (40 \%) variation near
the ``bridge'' magnitude, absent when an orthogonal cut is done.

Why not to use Kron magnitudes at all magnitude then, as many literature
works? The reason is written in the SExtractor manual: the Kron magnitude 
is measured in two different ways depending on the measured object 
radius: it is a true Kron magnitude for objects larger than a
radius threshold and an aperture magnitude for fainter objects. Such
a measure, and potentially all flip--flopping magnitudes (such
as ``auto" or ``best" mag) distorts the luminosity
distribution (i.e. the galaxy counts) near the 
``bridge" point.

The magnitude completeness is defined as the magnitude where objects start
to be lost because their measured central brightness is lower than the
detection  threshold (see Garilli, Maccagni \& Andreon 1999 and paper I for
details).   For AC\,114, the ($5\sigma$) limiting magnitude is
$K_s\sim20.3$ mag in a $4.4$ arcsec aperture. For simplicity and for excess
of caution, we consider here only $K_s<20.0$ mag objects.

Objects are classified according to their compactness, by using the SExtractor
stellar classifier. Almost the whole area of AC\,114 studied here has been
observed by the {\it Hubble Space Telescope} mosaic (Couch et al. 1998).
 Galaxies are resolved (i.e. not point--like) objects at the {\it HST}
resolution. The comparison of our ground-based classification and
{\it HST} images of the same objects
confirms the goodness of our ground--based star/galaxy
classification
because a few galaxies, out of hundreds, are misclassified.

\subsection{Comparison to literature photometry}

AC\,114 has been observed in the $K'$ band by Barger et al. (1996) and by
Stanford et al. (2002).  Stanford et al. (2002) measure aperture magnitudes
(in a 5 arcsec aperture).  Our Kron mag well agree with them, with no
photometric offset and a typical scatter of 0.2 mag. 

Barger et al. (1996) measure pseudo-total magnitudes on images taken with
an instrument having a large pixel size  (0.79 arcsec). Our magnitudes are
brighter than their by 0.18 mag, the offset being potentially due to their
quite large pixel size and worse seeing (between
1.1 and 1.7 arcsec).

All objects listed in Barger et al. (1996) or Stanford et al. (2002) are
present in our catalog, as expected because our images are much deeper.
Instead, several objects, brighter than the completeness magnitude of
Barger et al. (1996) or Stanford et al. (2002) are missing in
their catalogs.

\section{LF, statistical methods}

\subsection{Background}

We are here faced with the classical problem of determining two  extended (integral$>1$) density probability
function, one carrying the signal (the cluster LF) and the other being due to a  background
(background galaxy counts, BKG) from the observations of many individual events (the galaxies
luminosities), without knowledge of which event is the signal and which one is background.

Traditionally, the cluster luminosity function is computed as the difference
between galaxy counts in the cluster and control field directions (Zwicky
1957, Oemler 1974), i.e. after binning the events (galaxy magnitudes) in magnitude bins. 
In performing such a computation, 

1) galaxy counts are
binned in magnitude bins (of arbitrary width) and 

2) galaxy counts in the 
control field direction are subtracted from counts in the cluster direction
in order to obtain the cluster contribution alone. 

3) in order to estimate the error on the cluster LF, 
approximate Poisson errors, i.e. $\sqrt n$, and in
some cases over--Poisson ones due to large scale structure are added in
quadrature, under the (approximate) hypothesis that they are Gaussian
distributed. 

 Binning has several advantages:

\begin{itemize}

\item
It allows to ``see" how data are distributed (or better, to ``see" the data
distribution convolved by the binning function).

\item
It allows a quick analysis of the data.

\item
It allows to calculate the goodness of fit in a simple way, using a
$\chi^2$.

\item
It provides a correct result at large signal to noise.

\end{itemize}

The bin width is arbitrary, but recently Takeuchi (2000) suggests a
legitimate rule in the case of bins all of the same width: 
the Akaike's Information Criterion can be used for optimal
choosing the number of bins. However, when galaxy counts change by
three order of magnitudes, as usual in computing LFs, such an approach 
is optimal {\it on average} but far from the optimal in the (faint) bins
populated by thousand of galaxies or in the almost empty (bright) bins.

Indeed, it would be preferable to avoid any binning of the data
for the following
reasons:

\begin{itemize}

\item
No matter which amplitude bin is chosen, it tends to be too wide in crowded
regions and too narrow in low--populated regions.  Adaptive binning,
i.e. of variable width, depending on the local population is a possible
solution, that, however, shares the problems listed below, and requires
a more elaborate fitting algorithm (an appropriate convolution of
the fitting function).

\item
Negative LF (as well as background galaxy counts) makes no sense
(since both functions are positively defined), hence any determination
allowing the LF being negative has a dubious meaning.
Binning, coupled with background subtraction, may produce such 
occurrences: it may happen that,
because statistical fluctuations, the counts in the control field
direction are larger than the one in cluster direction, leaving a
negative number of galaxies, for a positively defined quantity.
Do the reader ever saw a negative number of galaxies?  Although
negative values are often consistent with zero, they cannot be simply
ignored or set to zero, otherwise a significant bias would
occur. For example, the integral over the LF, the cluster richness
and the luminosity density are systematically over-estimated.

\item
Binning frequently produces
error bars on $LF$ crossing the $LF=0$ line, considering
the possibility 
of a negative number of galaxies
(that the authors are still not ready to accept). 

\item
Binning
in high dimensions (here we have, for example, six to nine dimensions,
see eq. 3) 
makes the data sparse, no matter how large the sample is,
especially when the galaxy density change by three order of
magnitude from bright to faint magnitudes. As mentioned, low
populated regions are a problem for several reasons.

\item
Binning implicitly assumes that no change is occurring
inside the bin, and it occurs only at the bin boundaries (the idea
of continuity is lost in binning). For example, LFs are often measured in
redshift bins, assuming they do not evolve
inside the redshift bin, and then compared among them for 
looking a redshift evolution, that, according
to the logic of the people making such a comparison
occurs at the bin boundary only, and with ``quantum jumps" (see
Andreon 2004 for details). 

\item
Binning makes a rigorous statistical analysis
a nightmare: errors are not Gaussian distributed (when
the number of objects inside a bin is small, and in a multi
dimensional space there are always bins low populated),
linear last square fits (such as the $\chi^2$) badly fail
and give biased results when the number of objects inside
the bin is small (Wheaton et al. 1995). The latter work suggest
to fit ``one count at a time", id est not to bin at all. 
Furthermore, having observed $n_0$ the 68 \% confidence interval
is not given by $[n_0-\sqrt n_0, n_0+\sqrt n_0]$ when $n_0$ is small
(see, e.g., Gehrels 1986 or statistical textbooks)

\end{itemize}

We therefore opt for an unbinned fit of ``one galaxy at a time",
following the path put forward by Sandage, Tammann \& Yahil (1979, 
STY), where it was assumed no background, no evolution and no 
environment dependence to be present. 
In Lin et al. (1996) a monolithic (i.e. independent on luminosity) 
extension has been introduced under
the assumption of no background at all (i.e. the redshift of each
galaxy is known). Andreon (2004) remove the
monolithic evolution, allowing galaxies of different luminosity
to evolve by different amounts, still in absence of
background. 

In the present paper we allow
the presence of background galaxies, unrelated to the
cluster, i.e. we present how the LF can be computed when
the individual membership of galaxies is unknown. However,
we assume, as STY, a LF universality (i.e. a LF independent
on environment). The reason is mainly technical, not theoretical:
the formalism introduced below is easily extensible to such a case
(for example following the parametrization with
environment or redshift outlined in Andreon 2004),
but coding it is quite complex.

In order to account for observations of different quality
(dept, area, etc.)
a determination using several datasets (each one having bounds in 
magnitude or area) is allowed,
as in Efstathiou, Ellis, \& Peterson (1988, hereafter EEP). For example,
the used HDF--S
observations are actually left--censored at  (i.e. we have no
data at left of) $K_s=19.0$ and
right censored at  (i.e. we have no
data at right of)  $K_s=20.25$.

The method naturally converges to results obtained when
data are binned, when binning can be done, i.e. when
the number of objects for bin is large and
the Gaussian approximation occurs.

\subsection{Adopted approach}

\subsubsection{Using an extended likelihood and properly accounting for 
background}

Our approach is based on a single likelihood function, that accounts
simultaneously for all available data, cluster and control fields. The use
of the extended likelihood keep the normalization usually
lost in other methods. We don't require that the observed
background in the cluster line of sight is ``average" (or typical), but
only that it is drawn from the same parent distribution from which the
background in the control field is drawn.

Given $j$ datasets (say, cluster n. 1, cluster n. 2, ... field n. 1, field n. 2,
...) each composed of $N_j$ galaxies, 
we maximize the {\it extended} likelihood $\mathcal{L}$ given  by the formula:

\begin{equation}
\ln \mathcal{L} = \sum_{datasets \ j} ( \sum_{galaxies \ i} \ln p_i -s_j) 
\end{equation}

where: 

$p_i$ is the (extended, because integral is not 1) 
probability of the $i^{th}$ galaxy of the $j^{th}$
dataset to have $m_i$, i.e., $p_i=p(m_i)$

$s$ is the integral of function $\phi$ over the range
$[mag_{left,j},mag_{lim,j}]$, i.e. the expected number of
galaxies, given the model. In formula:

\begin{equation}
s = \int^{mag_{lim,j}}_{mag_{left,j}} \phi(m) dm 
\end{equation}

$mag_{lim,j}$ is the limiting magnitude of the $j$-th dataset,

$mag_{left,j}$ is the limiting magnitude at the bright end (in the
case of left--bounded mag values) of the $j$-th dataset. For example,
if in the sample $K<10$ galaxies are filtered out (because saturated, 
or because such galaxies would get trouble to the instrument by,
say, occupying a large fraction of the field of view), it will be
$mag_{left}=10$ for that sample.

$\phi$ is the sum of a power law (accounts for the background contribution) 
and a Schechter (1976) function:

\begin{equation}
p_i=\phi(m)=\delta_c
\Omega_j\phi^*10^{0.4(\alpha_z+1)(m^*-m)}e^{-10^{0.4(m^*-m)}}+
\end{equation}
$$ +\Omega_j 10^{a+b*(m-20)+c*(m-20)^2}$$ 

where $\delta_c=1$ for cluster datasets, $\delta_c=0$ for
the other datasets, 
$a,b,c$ describe the shape of the galaxy counts in the
reference field direction  and $\Omega_j$ is the studied
solid angle. The number ``20" is there
for numerical convenience. If galaxy counts have a more
complex magnitude distribution, more coefficients 
(or any other parametrization) can be 
used to describe the shape distributions. Analogously,
if the cluster LF is more complex than a Schechter function,
say a sum over the LFs of the individual morphological types (e.g. Andreon
1998), the Schechter function in eq. 3 can be replaced with
the reader favorite function without affecting the overall approach.

The above approach neglects the effect of large scale structure,
and it is justified when the variance due to large scale structure
is much lower than the Poissonian variance. For $K_s>12$ mag, and  
for a solid angle as small as one single 
SOFI field of view (about $20-25$ arcmin$^2$),
the variance due to large scale structure, computed according
to Huang et al. (1997), is less than 1\% of
the Poissonian variance and can be safely neglected.

The cluster LF is given by the Schechter (or favorite) function with 
parameters that maximize the likelihood.  Confidence contours may be
computed using the likelihood ratio theorem. The 68 \% and 95 \%
confidence contours  for two interesting parameters are computed from  $2
\Delta \ln \mathcal{L} = 2.3, 6.17$, respectively (Avni 1976; Wilks 1938,
1963, Cash 1979; Press et al. 1996). The 68 \% confidence interval for
a single parameter is computed using $2 \Delta \ln \mathcal{L} = 1$ (Avni
1976; Wilks 1938, 1963, Cash 1979; Press et al. 1996). We remind the
approximate nature of them and that some regularity conditions are
required (see Protassov et al. 2002 for astronomical
related  references). The large ($>1000$) number of galaxies and the
absence of borders near the best fit parameters guarantees that the
hypothesis on which the likelihood ratio  theorem is based are satisfied
for the data used in the present paper. Regularity conditions are not
always satisfied when dealing with the Butcher--Oemler effect (Andreon et al.
2005).

As goodness-of-fit we adopt the Persson's $\chi^2$ test, accurately
described in Sec 14.3 of Press et al. (1993) for Poissonian distributed
quantities.  The Persson's $\chi^2$ is, in the long run, $\chi^2$-distributed
with a number of degree of freedom, $\nu$, equal to the number of the bins
minus the number of parameters of the fit function. The test is applied on
galaxy counts, not on the difference of galaxy counts in the cluster and
control field directions. The goodness-of-fit estimation requires to bin the
data.

\subsubsection{How to find a global minimum}

The maximum of the likelihood can be found using simulated annealing methods
(e.g. Press et al. 1996), because the desired global maximum is often
hidden among many, poorer, local maxima in high dimensional spaces. For
larger dimension problems it is computationally more efficient to use Markov
Chain Monte Carlo (e.g. Dunkley et al. 2005).

Best fit parameters are determined all together at once:
we avoid the procedure used by other authors of  fitting
the control field counts, and, once found the best
fit parameters for the control field, switch to fit the 
the cluster counts by keeping background parameters 
fixed. The above procedure does not guaranteed 
to find the global minimum. 

Such a global fitting also accounts for a difference in the observed
value of background counts in cluster and reference field directions.

\subsection{Where we improve with respect to previous approaches}

In this section we summarize the differences between our 
approach and previous ones.

\subsubsection{STY}

STY and other maximum likelihood approaches don't deal with
the presence of a background, and hence cannot be used when the
individual membership of galaxies is unknown. 

It is well known that in the STY approach the normalization is lost
(i.e. $\phi^*$).  This situation is not typical
of maximum likelihood methods in general, and, in fact, the normalization
is kept in our approach
that also gives rigorous 68 \% confidence intervals, and this
is a good reason to adopt it.

\subsubsection{EEP}

The EEP method don't deal with
the presence of a background, and hence cannot be used when the
individual membership of galaxies is unknown. Furthermore,
EEP need to bin the data. Sec 3.1 explains why we dislike binning
the data.

\subsubsection{Wrong Poisson errors for small $n$}

As mentioned, LF determinations derived by binning the data
in magnitude bins and by computing the cluster contribution as
straightforward $cluster - field$ difference 
have error bars difficult to be computed, because 
for small $n$, the
68 \% confidence interval is not given by  $[n-\sqrt n ,n+\sqrt n]$
(e.g. Gehrels 1986), and the 68 \% confidence
interval on the
difference is not given by the quadrature sum of the 68 \% confidence
intervals of the two addenda.

Since we don't bin the data, we avoid to deal with those incorrect
expressions.

\subsubsection{Binning but forgetting to marginalize the model
over the bin}

Several LF methods bin the data in $mag$. Obviously,
the change does not occurs at the border bin.
One should, therefore, marginalize (integrate) the model LF over the
quantity binned. Such a rule is used in several papers for the "mag"
quantity (e.g. Paolillo et al. 2001),  but not systematically by all
authors. Said simply, some authors sincerely believe that inside the
bin there is only one "mag" and they compute errors as if this belief
is true. However, when describing the LF these authors don't
write that the LF is a sum of delta function, each one centered
at the center of the bin, but a smooth function, in logical
contradiction with having assumed a sum of delta functions.

\subsubsection{Forgetting $s$}

The $s$ term in the likelihood is required, as long as Poisson
fluctuations are allowed. If absent, or replaced by the observed
number of galaxies, Poisson fluctuations at each
$m$ are allowed, but Poisson fluctuations of the total number of
objects are not! 

In particular, neglecting $s$ in presence  of small signals (i.e. the
only occasion when statistics is actually required) is dangerous, in
the sense that even meaningless results can be found (for example, a
negative number of cluster galaxies), and usually leads to underestimating
the uncertainty on the parameters. Overlooking $s$ is quite
standard in the astronomical community, in the LF computation, in
recent detections of cluster of galaxies jointly using (Rosat) X--ray
photons and (SDSS) galaxy catalogs, etc.

Popesso et al. (2004) adopt a maximum likelihood method but they replace
$s$ with the number of  observed galaxies (see their eq. 4 and related
comments).  Their algorithm fails to find a reasonable best fit parameters in
several cases (look for $M^*=0.00$ values in their Table 2), the error on the
best fit parameter is found in some cases to be less than $0.005$ (for example
for RXCJ0747.0+4131), a precision never previously achieved not even with a 10
times larger sample, $M^*$ of a z=0.78 cluster (RXCJ1140.3+6609) can be
computed with good accuracy using about 50s  exposure at a 2.5m telescope,
when its brightest galaxies are marginally detected, if any. By replacing $s$
with the number of  observed galaxies may produce failures in finding 
reasonable values for the best fit parameters and may give strongly
underestimated uncertainties.  The $s$ term, prescribed by the extended
likelihood approach, does not allow similar situations to occur.

The latter work disagree with us in computing the LF from incomplete samples 
without accounting for incompleteness.

\subsubsection{Dissenting views}

The measure of the LF by using the statistical subtraction of background
has been criticized by Toft, Soucail, Hjorth (2003) that suggest an
alternative way to compute the LF ``without having to make uncertain
statistical corrections to account for foreground and background
contamination". A similar statement is repeated in Toft et al.
(2004), and in Blanton et al. (2005), because "background subtraction 
[is] an uncertain procedure".

First of all, it is unclear to us why the background subtraction is
uncertain. It is known with a degree of accuracy that
depends on the available data, as other experimental quantities.

Second, Toft, Soucail, Hjorth (2003) 
replace it with a photometric redshift selection plus
a correction for galaxies lost in  the selection. Such a correction
is uncertain, because it requires to know the distribution of
spectral types at the observed redshifts, and the spectral templates
at these redshifts. Both of them are unknown, at the difference
of background counts that are known, because computed in a control 
field.

Blanton et al. (2005) solution, instead, is to adopt a method (EEP)
which assumes that the LF is independent on environment (eq. 2.3 of EEP)
for a sample in which the dependence of the LF on environment is
flagrant (Fig 15 in Blanton et al. 2005).

Therefore, we cannot agree with their criticisms to the background
subtraction methods, and with their proposed solution.

The background subtraction method has been criticized by
Valotto et al. (2001), claiming that the presence of a background 
overdensity in the cluster line of sight favours the cluster detectability and
bias the slope of the luminosity function. The above occurs
often in their simulations, because ``many of the clusters 
found in two dimensions have no significant three-dimensional 
counterparts", as the authors claim. 
In nature, instead, most of (and perhaps all) 
the clusters whose LF is computed 
have a three-dimensional counterparts (i.e. when
spectroscopy is performed the cluster is confirmed), which simply
means that their simulations are not an accurate reproduction of
Nature.  Therefore, their criticism does not apply to actual data used
for the LF measure, but eventually applies to cases where
the cluster detection is doubful.  Furthermore,
the LF of a large sample of clusters in Paolillo et al. (2001), selected
in two dimensions by Abell (1958) and background subtracted
in the way criticized by Valotto et al. (2001),
is equal to the LF of another large sample of clusters 
(Garilli, Maccagni \& Andreon 1999) which is
x-ray selected and, according to Valotto et al. (2004), does
not suffer of the bias.
Therefore, the effect of the bias (if it exist) is negligible for the
data sets actually used.  
Finally, in the few cases when a cluster LF is determined by 
performing a spectroscopic survey deep enough to probe the
LF slope, the derived LF is
equal within the errors to the one derived by using a statistical 
background subtraction (e.g. for the Coma cluster:
Mobasher et al. 2003 vs Andreon \& Cuillandre 2002).

\begin{table}
\caption{Best fit parameters, errors and goodness of fit}
\begin{tabular}{llll}
\hline
& AC\,114 & AC\,118 & AC\,114 + AC\,118\\
\hline
$a$  &	  4.37095547	&  4.37931252 & 4.37309837\\
$b$  &	  0.303063065	& 0.312422931  & 0.305745333\\
$c$  &	  -0.0223323945 & -0.0203797854 & -0.0216595493\\
$\alpha$ &	$-1.30 \pm 0.07 $ & $-1.03 \pm 0.02$ & $-1.15 \pm 0.05$\\
$M^*$    &	$15.04 \pm 0.32 $ & $15.72 \pm 0.21$ & $ 15.43 \pm 0.14$\\
$\phi^*$ &	$6.3 \pm1.8 \ 10^3$ & $1.47 \pm 0.28 \ 10^4$ & 1.04 \& 1.03 $10^4$\\
$\chi^2$ & 39.0	(21.1) & 34.2 (12.3) & 49.0 (24.7)\\
$\nu$	 & 32 (15) & 33 (15) & 50 (24) \\
$P(\ge\chi^2)$ & 0.20 (0.15) & 0.40 (0.65) & 0.53 (0.40)\\
\hline
\end{tabular}
\hfill \break 
$a$, $b$ and $c$  describe the shape of galaxy counts (eq. 3), whereas       
$\alpha$, $M^*$ and $\phi^*$ describe the shape of the cluster LF (eq. 3).
Units: when inserted in eq. 3 $a$, $b$ and $c$ provide galaxy counts
in units of deg$^{-2}$. The latter are also the units of 
$\phi^*$. $M^*$ is given in mag units.
In the first three lines, there are more decimals than 
precision allows, to avoid truncation errors. 
The last three lines quote values including all (0.5 mag wide) bins
or, in parenthesis, excluding bins with less than 10 galaxies.
\hfill \break
\end{table}

\begin{figure}
\psfig{figure=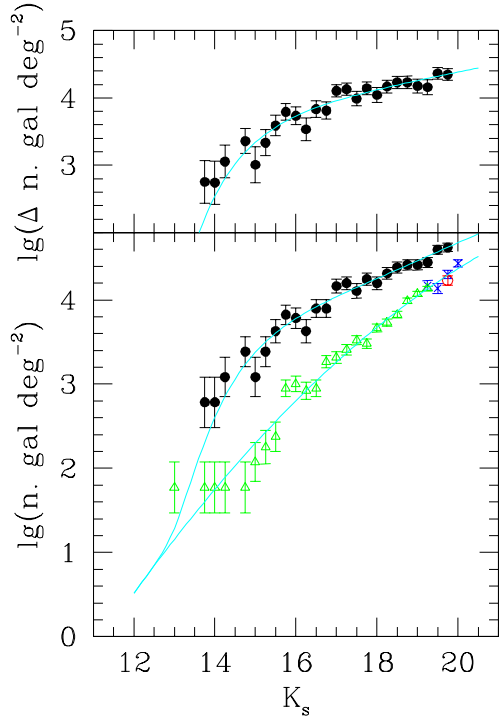,width=8truecm}
\caption[h]{Upper panel: The $K_s$ band LF of AC\,114.
Lower panel: galaxy counts in the AC\,114 line of sight (solid 
upper points, in black) and in the control field (lower points, colour- 
and type- coded:
green triangles=CDF--S, red empty circles= deep part of CDF--S, blue crosses= 
HDF--S). Note that
because crowding some points are to be seen on the plot. Incomplete bins,
i.e. that cover a magnitude interval not completely explored by the 
observations, are not plotted. 
The cyan lines are best joint fit to the control-field \& cluster line
of sight directions on unbinned data. Data are binned in the figure
for display purpose only, and computed as described in the text. 
Note that bins are 0.25 mag wide, half the usual bin width.}
\end{figure}

\begin{figure}
\psfig{figure=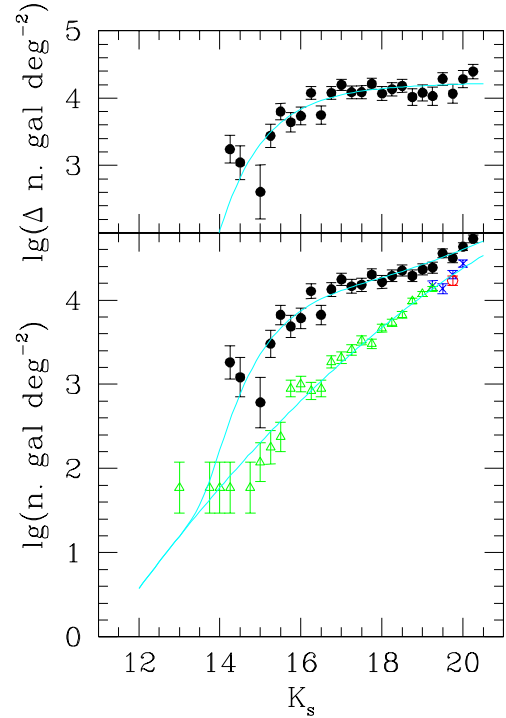,width=8truecm}
\caption[h]{As previous figure, but for AC\,118.}
\end{figure}

\section{Application of the method, the composite LF and discussion}

We apply the method to the data presented in Sec 2. Table 2 lists
best fit parameters and errors for Schechter parameters.

Figure 3 (for AC\,114) and 4 (for AC\,118) show the galaxy counts  in
the control field direction (lower points in the lower panel) and in
the cluster line of  sight (upper points in the lower panel), and a
joint fit to cluster and control field counts.  For display
purposes only, we show points and error bars computed with usual
recipes, although we make no use of them in our analysis (parameter
or errors determination). The fit  is performed on
the unbinned distributions, whereas we bin them for display purpose
only. According to astronomical standard practice, error bars 
in lower panels have a width given by $\sqrt n /\Omega_j$.
The points in the upper panels of Figures 3 and 4 
mark the algebraic difference between the 
galaxy counts in the cluster direction and the best fitting background
counts. When the difference is negative (at
$K<14$ mag, plus few points at fainter mag) the result cannot be plotted, 
because the scale requires a positive argument  for the logarithm. 
Error bars on upper panels of Figures 3 and 4 mark the
square root of the variances of the minuend and subtrahend.

The Schechter curve is instead the rigorous derivation of
the cluster LF, drawn
with the best fit parameters found on the (cluster+field, field) datasets. 
It is not a best fit to the 
$cluster-field$  difference, as
detailed in section 3.2.2 and it is positively defined at every
magnitude, at the variance of the above mentioned ``data" points. 
Nevertheless, the curves nicely describe the (approximatively computed)
cluster contribution, especially at large S/N, because here the two
approaches converge by definition. At $K_s=14.75$ mag for AC\,118
model predicts a number of galaxies similar to the data points of
adjacent bins, but the above mentioned difference takes an
unphysical value, the unpleased situation discussed in Sect. 3.1.

The fits are good, in the sense that the probability of finding a worser
$\chi^2$ is large (Table 2).

These LFs determinations function are among the deepest at the
studied (large) area ever measured (see Fig 10 in Paper I), 
to our best knowledge. We hope that our LF method makes them also
the most rigorously determined.

A question naturally arises. Are our improvement formally  correct but of
null importance? After all, the best fit model passes through the cluster
contribution, even if approximatively computed. So why one should bother
himself with all these apparently useless complications?

Our method don't produce puzzling results, and it is the appropriate choice
when puzzling results are found, for example when are observed:

- negative star formation rates, that, according
to the authors ``lack physical sense"  (Rojas et al. 2004); 

- negative flux densities (for some SCUBA sources, Smith et al. 2001,
their Table 1);

- clusters with negative blue fraction (Butcher \& Oemer 1984, 
their Fig 3);

- clusters with negative masses (at least in some magnitude/radial bin,
see e.g. Hansen et al. 2005, their figure 5, top-left panel).

Most of these (and other) puzzling results originates from not fully
accounting for the impact of a background and of its fluctuations 
in computing the quantity of
interest. Either the analysis is rigorous, and we are sure that the
result makes sense, or the correctness of the results cannot be
guaranteed.

Figure 5 shows 68 \% and 95 \% confidence contours on $m^*$ and
$\alpha$. 
With respect to confidence contours of AC\,118 computed in
Paper I, here confidence contours shrank by
a factor two because of the better determination of the
background counts, and moved by one (old) sigma, because the
new background counts do not longer show a minor excess, with respect
to a power law, at intermediate magnitudes.

\begin{figure}
\psfig{figure=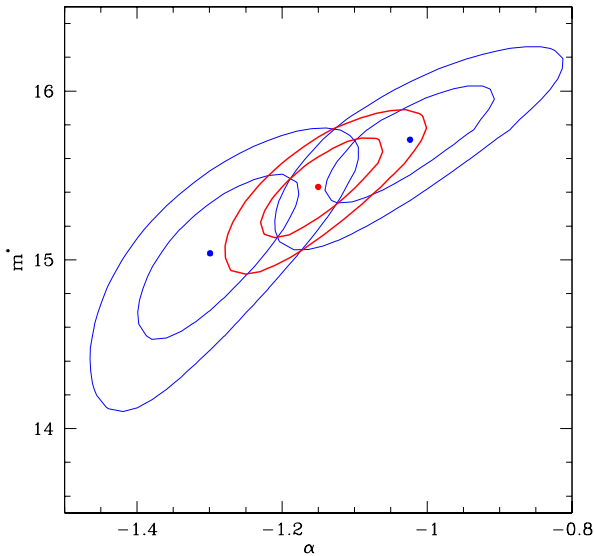,width=8truecm}
\caption[h]{68 \% and 95 \% confidence contours for $\alpha$
and $m^*$. The contours at the left, right and at the center  concern AC\,114, 
AC\,118 and the combined AC\,114+AC\,118 sample, respectively.}
\end{figure} 

Inspection of Figure 5 shows that two LFs are 
approximatively compatible at the 95 \% confidence level.
AC\,114 is, if any, steeper  
and brighter  than AC\,118, as comparison of 
Fig 2 and 3 also shows.
Such a difference is expected, given the dependence of the
best fit LF parameters on the environment, put forward in
paper I and Andreon (2002), and the observed difference in the
density distribution of the two clusters (compare Fig 1 with Fig 1
in Paper I). 

Although the inspection of the relative location of ($\alpha, M^*$) values and
contours is the standard astronomical way of comparing ($\alpha, M^*$) values
for different samples,  a rigorous comparison of the two LFs, however, should 
follow another path: first of all, Figure 5 shows that there are ($\alpha,
M^*$) values within both 95 \% confidence contours, but don't tell if these
values are achieved for the same values of the parameters not plotted in
Figure. For our LFs, the nuisance parameters are the background ($a,b,c$)
parameters. For field LFs the nuisance parameter $\phi^*$ is ``hidden'' and
the 95 \% confidence contours of the two compared LF may overlap, but for
incompatible $\phi^*$ values. Second, the simple comparison of Figure 5 may
incorrectly leave the impression of compatible LFs when instead the two LFs are
actually different. Consider, for example, the case of two LFs very different,
but computed for  a background known within a large uncertainty (leading to
large confidence  contours): the derived contours overlap each other, while a
correct comparison of the data (see below) will show the two LFs to be
different. Finally, and even in absence of a background (or any nuisance
parameter), what actually the figure shows is that there is a region of {\it
observed} values of the $\alpha, M^*$ plane (the region where confidence
contours cross)  that can be drawn from two different true values of $\alpha,
m^*$ at a given confidence, and {\it not} that a single pair of  $\alpha, M^*$
may generate two {\it observed} $\alpha, M^*$ at that confidence: confidence
contours give the probability of the data given the hypothesis and not vice
versa. By the way,  the contours for the two clusters are computed for
different hypothesis and both cannot be true at the same time (the two pairs
of best fit parameters are numerically different). 

In order to establish if the two LFs differ, we can ask ourself if 
a fit of both clusters with a single set of $\alpha, m^*$ values is much worser
that a fit with individual $\alpha, m^*$ values for each cluster. The likelihood
ratio test (LRT) allow such a comparison, and without any need to bin the data.
Our models to be compared are hierarchically nested\footnote{For example, 
the allowed parameter values of one model must be a subset of
those of the other model.} and regularity conditions
needed to use the LRT hold in our case (but see Protassov et al. 2002 for
a case where regularity conditions do not).
The likelihood ratio is computed by taking the ratio of the maximum value of the
likelihood function under the constraint of the null hypothesis (=one set of
$\alpha, m^*$ values) to the maximum with that constraint relaxed (=two sets of
$\alpha, m^*$ values). If the null hypothesis is true, then $2 \ \Delta \ln
\mathcal{L}$  (=twice the above ratio expressed in logarithm units) will be
asymptotically $\chi^2$-distributed with a number of degrees of freedom equal to
the difference in dimensionality of the two considered models. 

Therefore, we modify eq. 3 adding one more Schecther function, and we
fit at once all data, once keeping one single set of $m^*, \alpha$ values
for both clusters, and once leaving free $m^*, \alpha$ for each cluster
independently. In each fit, both clusters share the same set of
parameters describing the background, at variance of the fits discussed earlier in
this section, when
we did'nt constraint the parameters of the background to be the same in
the AC\,114 and AC\,118 fits.

We measure $2 \ \Delta \ln \mathcal{L} = 6.4$ for two (more) degrees of freedom.
Therefore, under the null hypothesis (the two LF having the same $m^*, \alpha$
parameters), there is about 5 \% probability to observe a larger likelihood ratio,
confirming the cursory inspection of AC\,114 and AC\,118 confidence 
contours previously mentioned. Such a
probability is not small enough to reject the null hypothesis that the two LF have
the same $m^*, \alpha$ parameters at high confidence, and we can therefore co--add
the data of the two clusters and compute the composite LF, which is, actually, the
likelihood under the null hypothesis just computed. The above path naturally solve
to the difficult procedure of average the LFs of the two clusters (or, more
generally two data sets), rigorously accounting for the error on the relative
normalization of the two LFs, often not even mentioned in astronomical papers.

Best fit
values for the combined data set (=AC\,114+AC\,118) are listed in Table 2 and
$m^*, \alpha$ confidence contours are shown in Figure 5. 
The fit is good, in the sense that the probability of finding a worser
$\chi^2$ is large (Table 2).

Although the use of a rigorous (and time consuming) test leads to the same
conclusion of a cursory inspection of AC\,114 and AC\,118 confidence contours, 
the former guarantees a correctness that the latter does not, and therefore
should be preferred.

\section{Summary and comments}

We presented a rigorous method to measure the luminosity function
in presence of background, extending previous methods to deal
with a more complicate case, and including neglected terms.
The approach does not suffer from logical inconsistences (or limitations) 
present in previous approaches and put on a sure foot claims of
providing errors with the correct coverage (i.e. that our errors
are 68 \% confidence intervals). We applied
the method to measure the LF of two clusters of galaxies,
using among the deepest and widest observations in $K_s$ band
ever done, producing one of the best determinations of the LF in
this band (and we hope one of the more rigorous ones, too). In passing
we show the bias of a flip--flopping definition of magnitude, and we
remind that several type of magnitudes are flip--flopping.
Several of our comments are clearly not specific to cluster
LFs and holds for the {\it field} LF too.

Distribution functions in presence of background (such as the
cluster LF in absence of a redshift survey, but also the H alpha
equivalent width distribution in presence of a continuum) should 

-- be fitted without removing the background contribution, 
adding instead a background term to the model,

-- be simultaneously fitted with the background distribution, 

-- use unbinned data, 

-- adopt the likelihood (not the conditional likelihood),

-- allow Poisson fluctuations of the whole sample (i.e. include the
$s$ term, as prescribed by the extended likelihood approach),

-- avoid the use of $\sqrt n$ in place of the 68 \%
confidence interval 

-- do not compute the 68 \% confidence interval by summing in
quadrature the 68 \% confidence intervals of the addenda.

Two sources of errors are negligible in our work, and therefore
neglected.
First, the error on the value of the input quantities, that in our case
are magnitudes, but in other papers are magnitudes and densities. With
the operated choices, magnitudes have negligible errors, and for this
reason we have neglected their impact on the LF parameters. 
If this condition does not
arise, one need to convolve the fitting function by the error
function, in the way described by Jeffreys (1938). Densities, instead,
usually have large errors, as large as 100 \% (for
example in some 2dF sub-samples, from quantities quoted in Croton et al.
2005), simply because densities are computed by counting a small
number of galaxies (e.g. as few as 1). $\Sigma_5$, a measure of
density derived from the distance of the 5$^{th}$ neighbour,
used in some recent
density estimates, has a $\pm55$ \% error.
The presence of large errors on the
input quantity further complicates a rigorous determination of the LF
and of his dependence on environment. Such a rigorous determination
has not yet been published, to our best knowledge. 

Second, we studied the whole galaxy population, and not a minority
population. In the latter case, errors in
the galaxy classification, even if coming infrequently, pollute the
minority population with objects coming from the main sample. Let us
consider an example: the LF of a population representing a tiny 
fraction, say 0.0003, as the fraction of local E+A galaxies.
If classification errors concern a fraction of 0.0003
of the whole galaxy population (a very small fraction, indeed), any
E+A sample studied is 50 \% contaminated by objects 
unrelated to the class aimed to study, and one should not be
surprised to ``discover" that the selected sample has a
LF similar to the one of the whole sample, being 50 \%
contaminated. Such contamination should be
accounted for in the LF computation, but it is usually not. Our 
approach of not subtracting the background from the data, but of
to add a background term to the model, accounts for the uncertainty
due to the mentioned contamination population.

\section*{Acknowledgments}

\label{lastpage}
\bsp

\end{document}